\documentclass[a4paper, amsfonts, superscriptaddress, amssymb, amsmath, reprint, showkeys, nofootinbib, twoside]{revtex4-1}

\usepackage[english]{babel}
\usepackage[utf8]{inputenc}
\usepackage{comment}
\usepackage[colorinlistoftodos, color=green!40, prependcaption]{todonotes}
\usepackage{amsthm}
\usepackage{mathtools}
\usepackage{physics}
\usepackage{xcolor}
\usepackage{graphicx}
\usepackage[left=23mm,right=13mm,top=35mm,columnsep=15pt]{geometry} 
\usepackage{adjustbox}
\usepackage{placeins}
\usepackage[T1]{fontenc}
\usepackage{lipsum}
\usepackage{csquotes}
\usepackage[pdftex, pdftitle={Article}, pdfauthor={Author}]{hyperref} 
\begin{document}
\title{Partial Information Rate Decomposition}

\author{Luca Faes}
    \email[Correspondence email address: ]{luca.faes@unipa.it}
    \affiliation{Department of Engineering, University of Palermo, Palermo, Italy}
    \affiliation{Faculty of Technical Sciences, University of Novi Sad, Serbia}
\author{Laura Sparacino}
    \affiliation{Department of Engineering, University of Palermo, Palermo, Italy}
\author{Gorana Mijatovic}
    \affiliation{Faculty of Technical Sciences, University of Novi Sad, Serbia}
\author{Yuri Antonacci}
    \affiliation{Department of Engineering, University of Palermo, Palermo, Italy}
\author{Leonardo Ricci}
    \affiliation{Department of Physics, University of Trento, Italy}
\author{Daniele Marinazzo}
    \affiliation{University of Ghent, Belgium}
\author{Sebastiano Stramaglia}
    \affiliation{University of Bari Aldo Moro and Istituto Nazionale
di Fisica Nucleare, Sezione di Bari, Italy}

\date{\today} 

\begin{abstract}
Partial Information Decomposition (PID) is a principled and flexible method to unveil complex high-order interactions in multi-unit network systems. Though being defined exclusively for random variables, PID is ubiquitously applied to multivariate time series taken as realizations of random processes with temporal statistical structure. Here, to overcome the incorrect depiction of high-order effects by PID schemes applied to dynamic networks, we introduce the framework of Partial Information Rate Decomposition (PIRD). PIRD is \textcolor{black}{first formalized applying lattice theory to decompose the information shared dynamically between a target random process and a set of source processes, and then implemented for Gaussian processes through a spectral expansion of information rates. The new framework is validated in simulated network systems and demonstrated in the practical analysis of} time series from large-scale climate oscillations.
\end{abstract}


\maketitle

The pursuit of assessing and disentangling complex many-body interactions in network systems composed by multiple connected units is crucial to data-driven research in many scientific fields.
In this context, partial information decomposition (PID) constitutes a comprehensive framework designed to understand how information is distributed in multivariate systems \cite{williams2010nonnegative}. The goal of PID is to decompose the information that a "target" random variable shares with a set of "source" variables into components highlighting how such an information is distributed among the sources: the \textit{unique} information that is exclusively available from each source, the \textit{redundant} information that can be obtained from at least two different sources, and the \textit{synergistic} information that is revealed only when multiple sources are considered simultaneously.
This framework has gained popularity as a main tool to assess high-order interdependencies from network datasets collected in several applicative fields of physics, engineering, biology, and artificial intelligence \cite{wibral2017partial,cang2020inferring,rosas2020reconciling,luppi2022synergistic,wollstadt2023rigorous,dissanayake2024quantifying}.

\begin{figure} [t!]
    \centering
    \includegraphics[scale= 0.88]{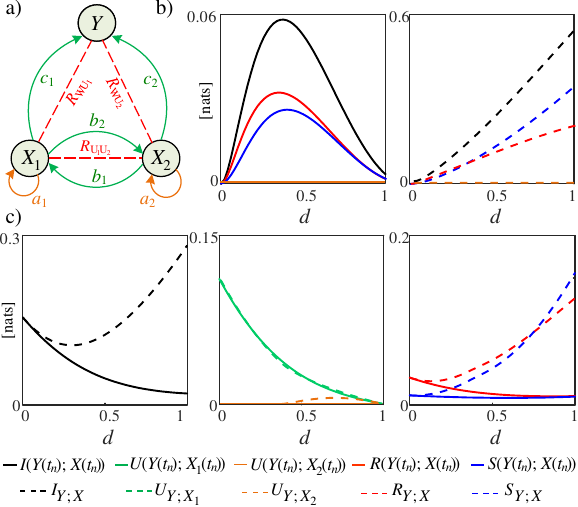}
    \caption{Decomposition of the multivariate information shared instantaneously and dynamically in a network Gaussian system with instantaneous (red) and time-lagged (internal: orange; causal: green) simulated interactions (a). In the absence of instantaneous effects, the zero-lag multivariate information is modulated non-trivially by time-lagged effects and is predominantly redundant (b, left), while the dynamic information reflects the strength of time-lagged effects and is predominantly synergistic (b, right). When a transition from pure instantaneous to pure time-lagged effects is simulated (c), PID (solid lines) fails in detecting the trend and nature of multivariate interactions, while PIRD reveals the expected emergence of dynamic coupling (dashed, black) and net synergy (dashed, blue vs. red).}
    \label{res_sim}
\end{figure}

Even though the PID is specifically defined for random variables, the data sequences over which it is computed are often multivariate time series that can only be regarded as realizations of random processes. In this case, under the implicit underlying assumption that the processes at hand are stationary and memoryless (i.e., composed by independent and identically distributed (i.i.d.) variables), the PID decomposes the information shared instantaneously by the processes.
However, the i.i.d. assumption is typically not tested in practice and is often violated in applications of information decomposition where the analyzed data exhibit temporal correlations \cite{kay2022comparison,varley2023information,varley2023multivariate}.
Here, we warn against the use of PID in processes exhibiting evident temporal statistical structure, showing that the presence of temporal correlations has a profound impact on the multivariate information shared at lag zero by multiple random processes (Fig. \ref{res_sim}).
On the other hand, applications of the PID to time-lagged variables selected ad-hoc from the processes (e.g.,\cite{luppi2022synergistic,koccillari2023behavioural,luppi2024synergistic}) are inherently focused on specific portions of the system dynamics, thus revealing the lack of general and comprehensive approaches to information decomposition that can provide a full account of the dynamical nature of random processes.
The present work fills this fundamental gap \textcolor{black}{first by formalizing the PID framework for networks of random processes through the use entropy rates as basic elements of information decomposition, and then developing a computationally reliable implementation of the approach based on a new definition of redundancy valid for information rates.}

We start considering the information-theoretic measure quantifying the overall degree of association between two discrete-time stationary random processes \textcolor{black}{$X=\{X(t_n)\}_{n \in \mathbb{Z}}$ and $Y=\{Y(t_n)\}_{n \in \mathbb{Z}}$}, which is the mutual information rate (MIR) \cite{duncan1970calculation} defined as
\begin{equation}
   I_{X;Y}=\lim_{n \to \infty} \frac{1}{n} I(X(t_1),\ldots,X(t_n);Y(t_1),\ldots,Y(t_n)), \label{MIR}
\end{equation}
where $I(\cdot;\cdot)$ denotes the mutual information (MI) between two random variables. 
Then, considering \textcolor{black}{$\textbf{X}$} as a vector of $N$ source processes,  $\textcolor{black}{\textbf{X}}=\{X_1,\ldots,X_N \}$, and denoting $Y$ as the scalar target process, we exploit the mathematical lattice structure defined for the PID \cite{williams2010nonnegative} and apply it to represent the set-theoretic intersection of multiple processes, so as to decompose the MIR as
\begin{equation}
    I_{\textcolor{black}{\textbf{X}};Y} = \sum_{\alpha \in \mathcal{A}} I^{\delta}_{\textcolor{black}{\textbf{X}}_{\alpha};Y},
    \label{PIRD}
\end{equation}
where $\textcolor{black}{\textbf{X}}_{\alpha} \subseteq \textcolor{black}{\textbf{X}}$ groups the source processes indexed by $\alpha$, and $\mathcal{A}$ is the collection of all subsets of sources such that no subset is a superset of any other (e.g., $\mathcal{A}=\{\{1\}\{2\},\{1\},\{2\},\{12\}\}$ if $N=2$) \cite{williams2010nonnegative}. Eq. (\ref{PIRD}) achieves a so-called \textit{partial information rate decomposition} (PIRD) whereby the MIR is expanded as the sum of contributions (information rate atoms $I^{\delta}_{\cdot;\cdot}$) identified by the lattice structure.
Moreover, the marginal MIR terms involving any individual source process $X_i$ are constructed additively by summing the information rate of the atoms positioned at the level $\{i\}$ and downwards in the lattice according to
\begin{equation}
    I_{X_i;Y}=\sum_{\beta \preceq \{i\}} I^{\delta}_{\textcolor{black}{\textbf{X}}_{\beta};Y}, i=1,\ldots,N
    \label{marginalMIR}
\end{equation}
where $\preceq$ identifies precedence based on the partial ordering imposed by the lattice structure \cite{williams2010nonnegative}. 
As happens with PID, the consistency equations (\ref{PIRD}) and (\ref{marginalMIR}) do not suffice to solve the PIRD problem because they provide a number of constraints lower than the number of information rate atoms to be computed (i.e., $N+1<|\mathcal{A}|$). Thus, to complete the PIRD it is necessary to define a so-called \textit{redundancy rate} function, here denoted as $I^{\cap}_{\cdot;\cdot}$, which generalizes the MIR over the lattice. The redundancy rate extends (\ref{marginalMIR}) to each atom $\alpha \in \mathcal{A}$, fulfilling 
\begin{equation}
    I^{\cap}_{\textcolor{black}{\textbf{X}}_{\alpha};Y}=\sum_{\beta \preceq \alpha} I^{\delta}_{\textcolor{black}{\textbf{X}}_{\beta};Y}.
    \label{RedRate}
\end{equation}
Then, once the redundancy rate is known, the information rate associated to all atoms can be retrieved via  M\"{o}bius inversion of  (\ref{RedRate}).
When $N=2$ source processes are considered, the PIRD atoms identify the redundant, unique and synergistic information rates (respectively, $R_{Y;\textcolor{black}{\textbf{X}}}\vcentcolon=I^{\delta}_{\textcolor{black}{\textbf{X}}_{\{1\}\{2\}};Y}$, $U_{Y;X_1}\vcentcolon=I^{\delta}_{X_{\{1\}};Y}$ and $U_{Y;X_2}\vcentcolon=I^{\delta}_{X_{\{2\}};Y}$, and $S_{Y;\textcolor{black}{\textbf{X}}}\vcentcolon=I^{\delta}_{\textcolor{black}{\textbf{X}}_{\{12\}};Y}$), while for $N>2$ such contributions may result from coarse-graining approaches that sum the information rate of multiple atoms \cite{rosas2020reconciling}.
Being explicitly defined for random processes, the PIRD defined by (\ref{PIRD},\ref{RedRate}) generalizes previous approaches applying PID to random variables arbitrarily selected from the processes: in the absence of temporal correlations, the PIRD reduces to the static PID decomposing the instantaneous information shared by the processes, \textcolor{black}{$I(X(t_n);Y(t_n))$}; in the case of strictly causal processes with the target not sending information to the sources, it reduces to the PID applied to the joint transfer entropy from all sources to target, \textcolor{black}{$T_{\textbf{X}\rightarrow Y}=I(\textbf{X}(t_{<n});Y(t_n)|Y(t_{<n}))$, where $\textbf{X}(t_{<n})=\lim_{k \to \infty}(\textbf{X}(t_{n-k}),\ldots,\textbf{X}(t_{n-1})$)} \cite{sparacino2025PIRDpaper}.

The operationalization of the function needed to capture the notion of redundancy rate can follow several different strategies proposed to define redundancy among random variables. These strategies differ depending on the philosophy followed to satisfy the desired properties (e.g., decision-, game-, information-theoretic), on the nature (continuous or discrete) of the analyzed variables, and on assumptions made about their distribution (e.g., Gaussian) \cite{williams2010nonnegative,barrett2015exploration,ince2017measuring,gutknecht2021bits,ehrlich2024partial}.
A popular approach is to define redundancy as the ensemble (time-domain) average of a given \textit{local} redundancy measure obtained elaborating properly computed pointwise MI values \cite{ince2017measuring,finn2018pointwise}.
Here we follow a conceptually similar approach, making it specifically tailored to random processes in that (i) it is applied to MIR instead of MI and (ii) it is implemented through a local representation in frequency rather than in time.
\textcolor{black}{Formally, we define the redundancy rate of the atom $\alpha=\{\alpha_1,\ldots,\alpha_J\}\in \mathcal{A}$ as}
\begin{equation}
    \textcolor{black}{I^{\cap}_{\textbf{X}_{\alpha};Y}:=}\frac{1}{2\pi}\int_{-\pi}^{\pi}i^{\cap}_{\textcolor{black}{\textbf{X}}_{\alpha};Y}(\omega) \mathrm{d}\omega,
    \label{RedRateIntegral}
\end{equation}
where $i^{\cap}_{\cdot;\cdot}(\omega)$ is the \textcolor{black}{pointwise (local)}
\textit{spectral redundancy rate} defined at each normalized angular frequency $\omega \in [-\pi,\pi]$ according to the minimum MI principle \cite{barrett2015exploration} applied to the spectral MIR terms $i_{\textcolor{black}{\textbf{X}}_{\alpha_j};Y}(\omega)$:
\begin{equation}
    \textcolor{black}{i^{\cap}_{\textbf{X}_{\alpha};Y}(\omega):=} \min_{j=1,\ldots,J} i_{\textcolor{black}{\textbf{X}}_{\alpha_j};Y}(\omega).
    \label{SpectralRedRate}
\end{equation}

\textcolor{black}{To compute the redundancy rate and solve the PIRD it is necessary to provide a proper spectral MIR function to use in (6). Here, we exploit the linear representation of multivariate Gaussian processes, for which} the spectral MIR between any two analyzed processes can be computed from the elements of the power spectral density of the joint process $\textcolor{black}{\textbf{S}}=\{\textcolor{black}{\textbf{X}},Y\}=\{X_1,\cdots X_N,Y\}$, defined as \textcolor{black}{the $(N+1)\times (N+1)$ matrix $\textbf{P}_{\textbf{S}}(\omega)=\mathfrak{F}\{\textbf{R}_\textbf{S}(k)\}$, with $\textbf{R}_\textbf{S}(k)=\mathbb{E}[\textbf{S}^{\intercal}(t_{n-k})\textbf{S}(t_n)]$} being the covariance of $\textcolor{black}{\textbf{S}}$ \cite{sparacino2025measuring}.
\textcolor{black}{Specifically, if $\textbf{S}$ is a jointly stationary Gaussian process, the MIR between the target $Y$ and the source  $\textcolor{black}{\textbf{X}}_{\alpha_j} \in \textbf{X}_{\alpha} \subseteq \textbf{X}$ can be expanded in the frequency domain as $I_{\textcolor{black}{\textbf{X}}_{\alpha_j};Y}=\frac{1}{2\pi}\int_{-\pi}^{\pi}i_{\textcolor{black}{\textbf{X}}_{\alpha_j};Y}(\omega)\mathrm{d}\omega$, where the spectral MIR function is \cite{chicharro2011spectral}
\begin{equation}
    i_{\textcolor{black}{\textbf{X}}_{\alpha_j};Y}(\omega)=\frac{1}{2}\log \frac{|\textcolor{black}{\textbf{P}}_{\textcolor{black}{\textbf{X}}_{\alpha_j}}(\omega)| P_{Y}(\omega)}{|\textcolor{black}{\textbf{P}}_{[\textcolor{black}{\textbf{X}}_{\alpha_j}Y]}(\omega)|}, \label{spectralMIR1} 
\end{equation}
being $|\cdot|$ the matrix determinant and $\textcolor{black}{\textbf{P}}_{[\textcolor{black}{\textbf{X}}_{\alpha_j}Y]}$ the sub-matrix of the full spectral density $\textbf{P}_\textbf{S}$ relevant to $\textcolor{black}{\textbf{X}}_{\alpha_j}$ and $Y$.}
For instance, in the case of two sources, the spectral redundancy rates obtained via (6) are  $i^{\cap}_{X_{\{k\}};Y}(\omega)=i_{X_k;Y}(\omega)$ for the unique atoms ($k=1,2$),  $i^{\cap}_{\textcolor{black}{\textbf{X}}_{\{1\}\{2\}};Y}(\omega)=\min \{i_{X_1;Y}(\omega),i_{X_2;Y}(\omega)\}$ for the redundant atom, and $i^{\cap}_{\textcolor{black}{\textbf{X}}_{\{12\}};Y}(\omega)=i_{X_1,X_2;Y}(\omega)$ for the synergistic atom, where the spectral MIR is computed via (7) respectively with $\alpha=\{k\}$, $\alpha=\{1\}\{2\}$, and $\alpha=\{12\}$.

\textcolor{black}{When implemented by means of the spectral MIR (7), the redundancy rate defined in (5,6) holds only for stationary Gaussian processes admitting a spectral representation \cite{chicharro2011spectral}. Moreover, due to the lack of additivity, the redundancy rate (5) differs from the minimum time-domain MIR; in \cite{sparacino2025PIRDpaper} we show that $I^{\cap}_{\textcolor{black}{\textbf{X}}_{\alpha};Y}\leq \min_{j=1,\ldots,J} I_{\textcolor{black}{\textbf{X}}_{\alpha_j};Y}$, denoting that applying the minimum MI principle at the local level in the spectral domain yields more conservative redundancy rates than working at the global level in the time domain. Nevertheless, the proposed redundancy rate preserves non-negativity, satisfies the main axioms of redundancy measures (i.e. symmetry, self-redundancy, and monotonicity) \cite{williams2010nonnegative}, and offers the interesting possibility to perform PIRD focusing on predefined frequency bands with practical meaning \cite{sparacino2025PIRDpaper}.}

\textcolor{black}{To validate PIRD in an illustrative example}, we consider a simple network system whose connections are mapped by the vector random process $\textcolor{black}{\textbf{S}}=\{X_1,X_2,Y\}$ with dynamics determined by the stochastic equations \textcolor{black}{
\begin{equation}
\begin{aligned}
    X_{1}(t_n)&=a_1X_{1}(t_{n-1})+b_1X_{2}(t_{n-1})+U_{1}(t_{n}), \\
    X_{2}(t_n)&=a_2X_2(t_{n-1})+b_2X_1(t_{n-1})+U_{2}(t_{n}), \\
    Y(t_n)&=c_1X_1(t_{n-1})+c_2X_2(t_{n-1})+W(t_n), \label{VARsimu}
\end{aligned}
\end{equation}}
where $W$ and $U_k$ are white and stationary innovation processes formed by i.i.d. standard Gaussian variables ($W(t_n),U_k(t_n)\sim \mathcal{N}(0,1)$, $k=1,2$) whose covariances $R_{U_1U_2}=\mathbb{E}[U_1(t_n)U_2(t_n)]$ and $R_{WU_k}=\mathbb{E}[W(t_n)U_k(t_n)]$ determine the strength of instantaneous (non-delayed) interactions (Fig. \ref{res_sim}a, red dashed lines).
The source processes $X_1$ and $X_2$ present internal dynamics modulated by the coefficients $a_1,a_2$ (orange arrows), and causal interactions are set between the sources and towards the target $Y$ via the coefficients $b_1,b_2$ and $c_1,c_2$ (green arrows); all these time-delayed interactions occur at lag 1.
\textcolor{black}{The balance between the strength of the causal mechanisms directed from sources to target ($X_1\rightarrow Y \leftarrow X_2$) or exerted between the sources ($X_1 \leftrightarrow X_2$) determines the prevalence of common target effects or common drive effects, which are known to induce respectively dominance of synergy or dominance of redundancy \cite{mijatovic2025network}}.

The system  is analyzed by computing both PID and PIRD \textcolor{black}{from the true values of the parameters imposed in (\ref{VARsimu}); details of the computation are reported in \cite{sparacino2025PIRDpaper}, while the code to simulate and analyze data is available at www.lucafaes.net/PIRD.html. The analysis is performed} as a function of a modulation  parameter $d \in [0,1]$ that acts on the strength of instantaneous and time-lagged effects.
First, we study (\ref{VARsimu}) in the absence of instantaneous effects ($R_{U_1U_2}=R_{WU_k}=0$), letting the strength of causal interactions to increase progressively ($c_1=c_2=d$), and the strength of internal dynamics to decrease progressively ($a_1=a_2=0.8(1-d)$), with fixed coupling between the sources ($b_1=b_2=0.1$). The analysis (Fig. \ref{res_sim}b) documents that the target shares information at lag zero with the sources as soon as time-lagged effects are set (\textcolor{black}{$I(Y(t_n);X(t_n))>0$} when $d>0$) even if instantaneous effects are kept to zero; this information does not exhibit a monotonic trend, and PID \textcolor{black}{incorrectly interprets it as prevalently redundant} \textcolor{black}{($R(Y(t_n);X(t_n))>S(Y(t_n);X(t_n))$)}. On the other hand, PIRD captures the increasing rate of information shared by the target and source processes, and documents the \textcolor{black}{expected emergence of net synergy consequent to the increasingly prevailing strength of the common target effect of $X_1$ and $X_2$ on $Y$}; the unique terms are null in this simulation where the two sources contribute equally to the target.
In the second simulation setting, the system is investigated reproducing a transition from purely instantaneous to purely time-lagged interactions among the processes, achieved decreasing progressively the strength of zero-lag effects ($R_{U_1U_2}=R_{WU_2}=0.25(1-d)$, $R_{WU_1}=0.5(1-d)$) while increasing the strength of internal dynamics and causal interactions ($a_1=a_2=0.2d$, $b_1=b_2=0.1d$, $c_1=c_2=0.6d$).
The results in Fig. \ref{res_sim}c confirm that the PIRD reduces to the PID in the absence of time-lagged effects ($d=0$), but the injection of temporal correlations into the system makes the two approaches to differ substantially. The information shared between \textcolor{black}{$Y(t_n)$ and $\{X_1(t_n),X_2(t_n)\}$} decreases progressively together with its unique, redundant and synergistic contributions, \textcolor{black}{preventing meaningful interpretations. On the other hand, PIRD allows describing expected trends, such as the initial decline of the overall information rate between $Y$ and $\{X_{1},X_{2}\}$ due to the decreasing unique contribution of $X_1$, and its subsequent increase due to the rise of both dynamic redundancy and synergy; net synergy emerges for $d>0.7$ when common target structures prevail over common drive ones.} 

As an application to real data, we consider an exemplary case study in climate science, i.e. the network of interactions among the most representative indices descriptive of El Niño and the Southern Oscillation (ENSO), a periodic fluctuation in the sea surface temperature and air pressure of the atmosphere overlying the equatorial Pacific Ocean constituting the most prominent interannual climate variability on Earth \cite{mcphaden2006enso}.
Since the exact initiating causes of an ENSO warm or cool events are not fully understood, it is important to analyze the statistical relation between its two main components, i.e. the East Central Tropical Pacific sea surface temperature anomaly (known as El Niño, NINO34) and the standardized difference in surface air pressure between Tahiti and Darwin (Southern Oscillation Index, SOI). These processes are dynamically related to several other indexes that represent large scale climate patterns \cite{chang2003tropical,silini2023assessing}, forming a 
network of causal interactions densely connected through several feedback loops of inter-annual variations \cite{silini2023assessing}.
Here, in addition to NINO34 and SOI we considered three more climate indices which can have a remarkable impact on ENSO as a result of high-order effects, i.e. TSA (Tropical Southern Atlantic Index), PDO (Pacific Decadal Oscillation), and NTA (North Tropical Atlantic) \cite{stramaglia2024disentangling}. \textcolor{black}{All indices are available at the NOAA website (https://psl.noaa.gov/data/climateindices/list/)} in the form of time series measured with a monthly sampling rate during the period 1950-2016 (792 data points).
The series were detrended and deseasonalized, and then analyzed under the assumption of Gaussianity so as to identify the terms relevant to the PID and PIRD formulations (the Gaussian assumption was adopted also in in \cite{silini2023assessing,stramaglia2024disentangling}). The series were analyzed in triplets, considering SOI as the target and all possible pairs of other series as sources; for each triplet, PIRD and zero-lag PID were computed respectively from the parameters of a vector autoregressive (VAR) model \cite{lutkepohl2005new} (least squares estimation, model order set according to the Akaike information criterion \cite{faes2012measuring}) and through simple linear regression (least squares estimation).
To assess the significance of temporal correlations, the PIRD was computed both on the original time series and on surrogate series for which the static MI was preserved (random shuffling of the samples of each series, with the same random permutation applied to all series).

\begin{figure}
    \centering
    \includegraphics{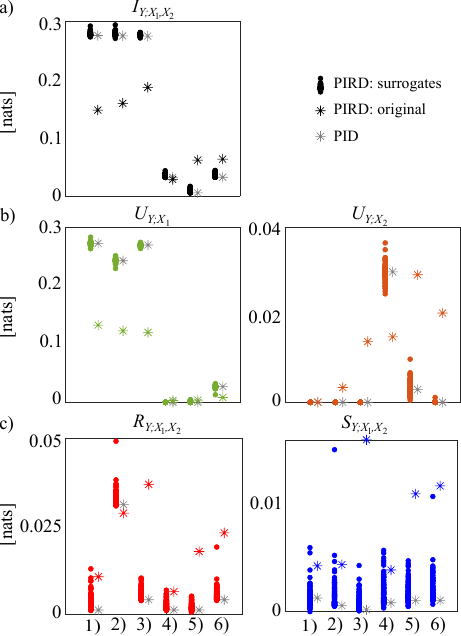}
    \caption{Decomposition of the multivariate information shared instantaneously and dynamically in six triplets of climate time series (target $Y$: SOI; sources $\{X_1, X_2\}$: 1) $\{NINO34,TSA\}$, 2) $\{NINO34,PDO\}$, 3) $\{NINO34,NTA\}$, 4) $\{TSA,PDO\}$, 5) $\{TSA,NTA\}$, 6) $\{PDO,NTA\}$).
    For each triplet, the multivariate information (a) is decomposed into unique (b) and redundant/synergistic (c) components via static PID (gray asterisks) and via the proposed dynamic PIRD (black/colored asterisks); PIRD is applied also on 100 surrogate triplets obtained via random shuffling preserving zero-lag correlations (circles).}
    \label{res_appl_climate}
\end{figure}

We found that the pairs of sources including NINO34 share the highest amounts of information with the target SOI (Fig. \ref{res_appl_climate}a), and that such information arises from a unique contribution of NINO34 (Fig. \ref{res_appl_climate}b); unique effects are noticeable also when considering NTA ($X_2$ in the triplets 3,5,6, Fig. \ref{res_appl_climate}b). The index NTA contributes significantly also to high-order interactions (both redundant and synergistic, triplets 3,5,6 in Fig. \ref{res_appl_climate}c). Crucially, the importance of NTA is elicited only using PIRD, \textcolor{black}{as synergistic PID values are negligible.} More generally, surrogate data demonstrate that temporal correlations play a fundamental role in information decomposition and cannot be disregarded; \textcolor{black}{the finding that PIRD measures can be both larger and smaller than the corresponding PID measures documents that time-lagged effects mix with each other and with instantaneous effects in a non-trivial way.}
These results confirm the main role played by El Niño temperature anomalies on athmospheric pressure and ENSO dynamics \cite{silini2023assessing, stramaglia2024disentangling}, \textcolor{black}{and suggest the importance of studying high-order interactions with a dynamic approach in this application. In fact, the redundant and synergistic dynamic effects emphasized only by PIRD among the two ENSO components and the NTA reveal the existence of wide climatological patterns which extend on a global scale involving the west southern pacific, east central pacific and north tropical atlantic climate modes.}

In summary, the proposed framework offers a novel comprehensive method to parcel out the information shared by dynamic network systems into amounts quantifying unique and high-order effects. We posit that this framework generalizes and unifies previous PID theories applied to multivariate time series, establishing a new approach for the analysis of complex networks that reconciles information-theoretic and dynamical systems perspectives. Indeed, as it explicitly takes the full temporal statistical structure of random processes into account, PIRD overcomes the intrinsic limitation of PID of being unable to deal with temporally correlated variables.
\textcolor{black}{The proposed spectral formulation of PIRD is valid for Gaussian systems and rests on the degree to which the system at hand can be described exhaustively by spectral analysis or linear models. This restriction does not invalidate the analysis of nonlinear systems, as stationary random processes are guaranteed to have a linear (albeit infinite-order) representation even if the generating mechanism of the process is nonlinear \cite{hannan1979statistical}. Moreover,} the formulation based on redundancy rates defined in the frequency domain allows to decompose dynamic information considering predetermined oscillations with specific meaning for the analyzed dynamic network \cite{sparacino2025PIRDpaper}. We expect that these properties, together with the algorithmic reliability and low computational demand of the implementation proposed here for PIRD, will open new avenues for the analysis of real-world networks with dynamic oscillatory behavior.

\begin{acknowledgments}
This work was supported by the project “HONEST - High-Order Dynamical Networks in Computational Neuroscience and Physiology: an Information-Theoretic Framework”, Italian Ministry of University and Research (funded by MUR, PRIN 2022, code 2022YMHNPY, CUP: B53D23003020006).
\end{acknowledgments}

\bibliography{biblio_letter.bib}

\begin{thebibliography}{29}%
\makeatletter
\providecommand \@ifxundefined [1]{%
 \@ifx{#1\undefined}
}%
\providecommand \@ifnum [1]{%
 \ifnum #1\expandafter \@firstoftwo
 \else \expandafter \@secondoftwo
 \fi
}%
\providecommand \@ifx [1]{%
 \ifx #1\expandafter \@firstoftwo
 \else \expandafter \@secondoftwo
 \fi
}%
\providecommand \natexlab [1]{#1}%
\providecommand \enquote  [1]{``#1''}%
\providecommand \bibnamefont  [1]{#1}%
\providecommand \bibfnamefont [1]{#1}%
\providecommand \citenamefont [1]{#1}%
\providecommand \href@noop [0]{\@secondoftwo}%
\providecommand \href [0]{\begingroup \@sanitize@url \@href}%
\providecommand \@href[1]{\@@startlink{#1}\@@href}%
\providecommand \@@href[1]{\endgroup#1\@@endlink}%
\providecommand \@sanitize@url [0]{\catcode `\\12\catcode `\$12\catcode `\&12\catcode `\#12\catcode `\^12\catcode `\_12\catcode `\%12\relax}%
\providecommand \@@startlink[1]{}%
\providecommand \@@endlink[0]{}%
\providecommand \url  [0]{\begingroup\@sanitize@url \@url }%
\providecommand \@url [1]{\endgroup\@href {#1}{\urlprefix }}%
\providecommand \urlprefix  [0]{URL }%
\providecommand \Eprint [0]{\href }%
\providecommand \doibase [0]{http://dx.doi.org/}%
\providecommand \selectlanguage [0]{\@gobble}%
\providecommand \bibinfo  [0]{\@secondoftwo}%
\providecommand \bibfield  [0]{\@secondoftwo}%
\providecommand \translation [1]{[#1]}%
\providecommand \BibitemOpen [0]{}%
\providecommand \bibitemStop [0]{}%
\providecommand \bibitemNoStop [0]{.\EOS\space}%
\providecommand \EOS [0]{\spacefactor3000\relax}%
\providecommand \BibitemShut  [1]{\csname bibitem#1\endcsname}%
\let\auto@bib@innerbib\@empty
\bibitem [{\citenamefont {Williams}\ and\ \citenamefont {Beer}(2010)}]{williams2010nonnegative}%
  \BibitemOpen
  \bibfield  {author} {\bibinfo {author} {\bibfnamefont {P.~L.}\ \bibnamefont {Williams}}\ and\ \bibinfo {author} {\bibfnamefont {R.~D.}\ \bibnamefont {Beer}},\ }\href@noop {} {\bibfield  {journal} {\bibinfo  {journal} {arXiv preprint arXiv:1004.2515}\ } (\bibinfo {year} {2010})}\BibitemShut {NoStop}%
\bibitem [{\citenamefont {Wibral}\ \emph {et~al.}(2017)\citenamefont {Wibral}, \citenamefont {Priesemann}, \citenamefont {Kay}, \citenamefont {Lizier},\ and\ \citenamefont {Phillips}}]{wibral2017partial}%
  \BibitemOpen
  \bibfield  {author} {\bibinfo {author} {\bibfnamefont {M.}~\bibnamefont {Wibral}}, \bibinfo {author} {\bibfnamefont {V.}~\bibnamefont {Priesemann}}, \bibinfo {author} {\bibfnamefont {J.~W.}\ \bibnamefont {Kay}}, \bibinfo {author} {\bibfnamefont {J.~T.}\ \bibnamefont {Lizier}}, \ and\ \bibinfo {author} {\bibfnamefont {W.~A.}\ \bibnamefont {Phillips}},\ }\href@noop {} {\bibfield  {journal} {\bibinfo  {journal} {Brain and cognition}\ }\textbf {\bibinfo {volume} {112}},\ \bibinfo {pages} {25} (\bibinfo {year} {2017})}\BibitemShut {NoStop}%
\bibitem [{\citenamefont {Cang}\ and\ \citenamefont {Nie}(2020)}]{cang2020inferring}%
  \BibitemOpen
  \bibfield  {author} {\bibinfo {author} {\bibfnamefont {Z.}~\bibnamefont {Cang}}\ and\ \bibinfo {author} {\bibfnamefont {Q.}~\bibnamefont {Nie}},\ }\href@noop {} {\bibfield  {journal} {\bibinfo  {journal} {Nature communications}\ }\textbf {\bibinfo {volume} {11}},\ \bibinfo {pages} {2084} (\bibinfo {year} {2020})}\BibitemShut {NoStop}%
\bibitem [{\citenamefont {Rosas}\ \emph {et~al.}(2020)\citenamefont {Rosas}, \citenamefont {Mediano}, \citenamefont {Jensen}, \citenamefont {Seth}, \citenamefont {Barrett}, \citenamefont {Carhart-Harris},\ and\ \citenamefont {Bor}}]{rosas2020reconciling}%
  \BibitemOpen
  \bibfield  {author} {\bibinfo {author} {\bibfnamefont {F.~E.}\ \bibnamefont {Rosas}}, \bibinfo {author} {\bibfnamefont {P.~A.}\ \bibnamefont {Mediano}}, \bibinfo {author} {\bibfnamefont {H.~J.}\ \bibnamefont {Jensen}}, \bibinfo {author} {\bibfnamefont {A.~K.}\ \bibnamefont {Seth}}, \bibinfo {author} {\bibfnamefont {A.~B.}\ \bibnamefont {Barrett}}, \bibinfo {author} {\bibfnamefont {R.~L.}\ \bibnamefont {Carhart-Harris}}, \ and\ \bibinfo {author} {\bibfnamefont {D.}~\bibnamefont {Bor}},\ }\href@noop {} {\bibfield  {journal} {\bibinfo  {journal} {PLoS computational biology}\ }\textbf {\bibinfo {volume} {16}},\ \bibinfo {pages} {e1008289} (\bibinfo {year} {2020})}\BibitemShut {NoStop}%
\bibitem [{\citenamefont {Luppi}\ \emph {et~al.}(2022)\citenamefont {Luppi}, \citenamefont {Mediano}, \citenamefont {Rosas}, \citenamefont {Holland}, \citenamefont {Fryer}, \citenamefont {O’Brien}, \citenamefont {Rowe}, \citenamefont {Menon}, \citenamefont {Bor},\ and\ \citenamefont {Stamatakis}}]{luppi2022synergistic}%
  \BibitemOpen
  \bibfield  {author} {\bibinfo {author} {\bibfnamefont {A.~I.}\ \bibnamefont {Luppi}}, \bibinfo {author} {\bibfnamefont {P.~A.}\ \bibnamefont {Mediano}}, \bibinfo {author} {\bibfnamefont {F.~E.}\ \bibnamefont {Rosas}}, \bibinfo {author} {\bibfnamefont {N.}~\bibnamefont {Holland}}, \bibinfo {author} {\bibfnamefont {T.~D.}\ \bibnamefont {Fryer}}, \bibinfo {author} {\bibfnamefont {J.~T.}\ \bibnamefont {O’Brien}}, \bibinfo {author} {\bibfnamefont {J.~B.}\ \bibnamefont {Rowe}}, \bibinfo {author} {\bibfnamefont {D.~K.}\ \bibnamefont {Menon}}, \bibinfo {author} {\bibfnamefont {D.}~\bibnamefont {Bor}}, \ and\ \bibinfo {author} {\bibfnamefont {E.~A.}\ \bibnamefont {Stamatakis}},\ }\href@noop {} {\bibfield  {journal} {\bibinfo  {journal} {Nature Neuroscience}\ }\textbf {\bibinfo {volume} {25}},\ \bibinfo {pages} {771} (\bibinfo {year} {2022})}\BibitemShut {NoStop}%
\bibitem [{\citenamefont {Wollstadt}\ \emph {et~al.}(2023)\citenamefont {Wollstadt}, \citenamefont {Schmitt},\ and\ \citenamefont {Wibral}}]{wollstadt2023rigorous}%
  \BibitemOpen
  \bibfield  {author} {\bibinfo {author} {\bibfnamefont {P.}~\bibnamefont {Wollstadt}}, \bibinfo {author} {\bibfnamefont {S.}~\bibnamefont {Schmitt}}, \ and\ \bibinfo {author} {\bibfnamefont {M.}~\bibnamefont {Wibral}},\ }\href@noop {} {\bibfield  {journal} {\bibinfo  {journal} {J. Mach. Learn. Res.}\ }\textbf {\bibinfo {volume} {24}},\ \bibinfo {pages} {1} (\bibinfo {year} {2023})}\BibitemShut {NoStop}%
\bibitem [{\citenamefont {Dissanayake}\ \emph {et~al.}(2024)\citenamefont {Dissanayake}, \citenamefont {Hamman}, \citenamefont {Halder}, \citenamefont {Sucholutsky}, \citenamefont {Zhang},\ and\ \citenamefont {Dutta}}]{dissanayake2024quantifying}%
  \BibitemOpen
  \bibfield  {author} {\bibinfo {author} {\bibfnamefont {P.}~\bibnamefont {Dissanayake}}, \bibinfo {author} {\bibfnamefont {F.}~\bibnamefont {Hamman}}, \bibinfo {author} {\bibfnamefont {B.}~\bibnamefont {Halder}}, \bibinfo {author} {\bibfnamefont {I.}~\bibnamefont {Sucholutsky}}, \bibinfo {author} {\bibfnamefont {Q.}~\bibnamefont {Zhang}}, \ and\ \bibinfo {author} {\bibfnamefont {S.}~\bibnamefont {Dutta}},\ }\href@noop {} {\bibfield  {journal} {\bibinfo  {journal} {arXiv preprint arXiv:2411.07483}\ } (\bibinfo {year} {2024})}\BibitemShut {NoStop}%
\bibitem [{\citenamefont {Kay}\ \emph {et~al.}(2022)\citenamefont {Kay}, \citenamefont {Schulz},\ and\ \citenamefont {Phillips}}]{kay2022comparison}%
  \BibitemOpen
  \bibfield  {author} {\bibinfo {author} {\bibfnamefont {J.~W.}\ \bibnamefont {Kay}}, \bibinfo {author} {\bibfnamefont {J.~M.}\ \bibnamefont {Schulz}}, \ and\ \bibinfo {author} {\bibfnamefont {W.~A.}\ \bibnamefont {Phillips}},\ }\href@noop {} {\bibfield  {journal} {\bibinfo  {journal} {Entropy}\ }\textbf {\bibinfo {volume} {24}},\ \bibinfo {pages} {1021} (\bibinfo {year} {2022})}\BibitemShut {NoStop}%
\bibitem [{\citenamefont {Varley}\ \emph {et~al.}(2023{\natexlab{a}})\citenamefont {Varley}, \citenamefont {Sporns}, \citenamefont {Schaffelhofer}, \citenamefont {Scherberger},\ and\ \citenamefont {Dann}}]{varley2023information}%
  \BibitemOpen
  \bibfield  {author} {\bibinfo {author} {\bibfnamefont {T.~F.}\ \bibnamefont {Varley}}, \bibinfo {author} {\bibfnamefont {O.}~\bibnamefont {Sporns}}, \bibinfo {author} {\bibfnamefont {S.}~\bibnamefont {Schaffelhofer}}, \bibinfo {author} {\bibfnamefont {H.}~\bibnamefont {Scherberger}}, \ and\ \bibinfo {author} {\bibfnamefont {B.}~\bibnamefont {Dann}},\ }\href@noop {} {\bibfield  {journal} {\bibinfo  {journal} {Proceedings of the National Academy of Sciences}\ }\textbf {\bibinfo {volume} {120}},\ \bibinfo {pages} {e2207677120} (\bibinfo {year} {2023}{\natexlab{a}})}\BibitemShut {NoStop}%
\bibitem [{\citenamefont {Varley}\ \emph {et~al.}(2023{\natexlab{b}})\citenamefont {Varley}, \citenamefont {Pope}, \citenamefont {Faskowitz},\ and\ \citenamefont {Sporns}}]{varley2023multivariate}%
  \BibitemOpen
  \bibfield  {author} {\bibinfo {author} {\bibfnamefont {T.~F.}\ \bibnamefont {Varley}}, \bibinfo {author} {\bibfnamefont {M.}~\bibnamefont {Pope}}, \bibinfo {author} {\bibfnamefont {J.}~\bibnamefont {Faskowitz}}, \ and\ \bibinfo {author} {\bibfnamefont {O.}~\bibnamefont {Sporns}},\ }\href@noop {} {\bibfield  {journal} {\bibinfo  {journal} {Communications biology}\ }\textbf {\bibinfo {volume} {6}},\ \bibinfo {pages} {451} (\bibinfo {year} {2023}{\natexlab{b}})}\BibitemShut {NoStop}%
\bibitem [{\citenamefont {Ko{\c{c}}illari}\ \emph {et~al.}(2023)\citenamefont {Ko{\c{c}}illari}, \citenamefont {Celotto}, \citenamefont {Francis}, \citenamefont {Mukherjee}, \citenamefont {Babadi}, \citenamefont {Kanold},\ and\ \citenamefont {Panzeri}}]{koccillari2023behavioural}%
  \BibitemOpen
  \bibfield  {author} {\bibinfo {author} {\bibfnamefont {L.}~\bibnamefont {Ko{\c{c}}illari}}, \bibinfo {author} {\bibfnamefont {M.}~\bibnamefont {Celotto}}, \bibinfo {author} {\bibfnamefont {N.~A.}\ \bibnamefont {Francis}}, \bibinfo {author} {\bibfnamefont {S.}~\bibnamefont {Mukherjee}}, \bibinfo {author} {\bibfnamefont {B.}~\bibnamefont {Babadi}}, \bibinfo {author} {\bibfnamefont {P.~O.}\ \bibnamefont {Kanold}}, \ and\ \bibinfo {author} {\bibfnamefont {S.}~\bibnamefont {Panzeri}},\ }\href@noop {} {\bibfield  {journal} {\bibinfo  {journal} {Brain Informatics}\ }\textbf {\bibinfo {volume} {10}},\ \bibinfo {pages} {34} (\bibinfo {year} {2023})}\BibitemShut {NoStop}%
\bibitem [{\citenamefont {Luppi}\ \emph {et~al.}(2024)\citenamefont {Luppi}, \citenamefont {Mediano}, \citenamefont {Rosas}, \citenamefont {Allanson}, \citenamefont {Pickard}, \citenamefont {Carhart-Harris}, \citenamefont {Williams}, \citenamefont {Craig}, \citenamefont {Finoia}, \citenamefont {Owen} \emph {et~al.}}]{luppi2024synergistic}%
  \BibitemOpen
  \bibfield  {author} {\bibinfo {author} {\bibfnamefont {A.~I.}\ \bibnamefont {Luppi}}, \bibinfo {author} {\bibfnamefont {P.~A.}\ \bibnamefont {Mediano}}, \bibinfo {author} {\bibfnamefont {F.~E.}\ \bibnamefont {Rosas}}, \bibinfo {author} {\bibfnamefont {J.}~\bibnamefont {Allanson}}, \bibinfo {author} {\bibfnamefont {J.}~\bibnamefont {Pickard}}, \bibinfo {author} {\bibfnamefont {R.~L.}\ \bibnamefont {Carhart-Harris}}, \bibinfo {author} {\bibfnamefont {G.~B.}\ \bibnamefont {Williams}}, \bibinfo {author} {\bibfnamefont {M.~M.}\ \bibnamefont {Craig}}, \bibinfo {author} {\bibfnamefont {P.}~\bibnamefont {Finoia}}, \bibinfo {author} {\bibfnamefont {A.~M.}\ \bibnamefont {Owen}},  \emph {et~al.},\ }\href@noop {} {\bibfield  {journal} {\bibinfo  {journal} {Elife}\ }\textbf {\bibinfo {volume} {12}},\ \bibinfo {pages} {RP88173} (\bibinfo {year} {2024})}\BibitemShut {NoStop}%
\bibitem [{\citenamefont {Duncan}(1970)}]{duncan1970calculation}%
  \BibitemOpen
  \bibfield  {author} {\bibinfo {author} {\bibfnamefont {T.~E.}\ \bibnamefont {Duncan}},\ }\href@noop {} {\bibfield  {journal} {\bibinfo  {journal} {SIAM Journal on Applied Mathematics}\ }\textbf {\bibinfo {volume} {19}},\ \bibinfo {pages} {215} (\bibinfo {year} {1970})}\BibitemShut {NoStop}%
\bibitem [{\citenamefont {Sparacino}\ \emph {et~al.}(2025{\natexlab{a}})\citenamefont {Sparacino}, \citenamefont {Mijatovic}, \citenamefont {Antonacci}, \citenamefont {Ricci}, \citenamefont {Marinazzo}, \citenamefont {Stramaglia},\ and\ \citenamefont {Faes}}]{sparacino2025PIRDpaper}%
  \BibitemOpen
  \bibfield  {author} {\bibinfo {author} {\bibfnamefont {L.}~\bibnamefont {Sparacino}}, \bibinfo {author} {\bibfnamefont {G.}~\bibnamefont {Mijatovic}}, \bibinfo {author} {\bibfnamefont {Y.}~\bibnamefont {Antonacci}}, \bibinfo {author} {\bibfnamefont {L.}~\bibnamefont {Ricci}}, \bibinfo {author} {\bibfnamefont {D.}~\bibnamefont {Marinazzo}}, \bibinfo {author} {\bibfnamefont {S.}~\bibnamefont {Stramaglia}}, \ and\ \bibinfo {author} {\bibfnamefont {L.}~\bibnamefont {Faes}},\ }\href@noop {} {\bibfield  {journal} {\bibinfo  {journal} {arXiv preprint arXiv:2502.04555}\ } (\bibinfo {year} {2025}{\natexlab{a}})}\BibitemShut {NoStop}%
\bibitem [{\citenamefont {Barrett}(2015)}]{barrett2015exploration}%
  \BibitemOpen
  \bibfield  {author} {\bibinfo {author} {\bibfnamefont {A.~B.}\ \bibnamefont {Barrett}},\ }\href@noop {} {\bibfield  {journal} {\bibinfo  {journal} {Physical Review E}\ }\textbf {\bibinfo {volume} {91}},\ \bibinfo {pages} {052802} (\bibinfo {year} {2015})}\BibitemShut {NoStop}%
\bibitem [{\citenamefont {Ince}(2017)}]{ince2017measuring}%
  \BibitemOpen
  \bibfield  {author} {\bibinfo {author} {\bibfnamefont {R.~A.}\ \bibnamefont {Ince}},\ }\href@noop {} {\bibfield  {journal} {\bibinfo  {journal} {Entropy}\ }\textbf {\bibinfo {volume} {19}},\ \bibinfo {pages} {318} (\bibinfo {year} {2017})}\BibitemShut {NoStop}%
\bibitem [{\citenamefont {Gutknecht}\ \emph {et~al.}(2021)\citenamefont {Gutknecht}, \citenamefont {Wibral},\ and\ \citenamefont {Makkeh}}]{gutknecht2021bits}%
  \BibitemOpen
  \bibfield  {author} {\bibinfo {author} {\bibfnamefont {A.~J.}\ \bibnamefont {Gutknecht}}, \bibinfo {author} {\bibfnamefont {M.}~\bibnamefont {Wibral}}, \ and\ \bibinfo {author} {\bibfnamefont {A.}~\bibnamefont {Makkeh}},\ }\href@noop {} {\bibfield  {journal} {\bibinfo  {journal} {Proceedings of the Royal Society A}\ }\textbf {\bibinfo {volume} {477}},\ \bibinfo {pages} {20210110} (\bibinfo {year} {2021})}\BibitemShut {NoStop}%
\bibitem [{\citenamefont {Ehrlich}\ \emph {et~al.}(2024)\citenamefont {Ehrlich}, \citenamefont {Schick-Poland}, \citenamefont {Makkeh}, \citenamefont {Lanfermann}, \citenamefont {Wollstadt},\ and\ \citenamefont {Wibral}}]{ehrlich2024partial}%
  \BibitemOpen
  \bibfield  {author} {\bibinfo {author} {\bibfnamefont {D.~A.}\ \bibnamefont {Ehrlich}}, \bibinfo {author} {\bibfnamefont {K.}~\bibnamefont {Schick-Poland}}, \bibinfo {author} {\bibfnamefont {A.}~\bibnamefont {Makkeh}}, \bibinfo {author} {\bibfnamefont {F.}~\bibnamefont {Lanfermann}}, \bibinfo {author} {\bibfnamefont {P.}~\bibnamefont {Wollstadt}}, \ and\ \bibinfo {author} {\bibfnamefont {M.}~\bibnamefont {Wibral}},\ }\href@noop {} {\bibfield  {journal} {\bibinfo  {journal} {Physical Review E}\ }\textbf {\bibinfo {volume} {110}},\ \bibinfo {pages} {014115} (\bibinfo {year} {2024})}\BibitemShut {NoStop}%
\bibitem [{\citenamefont {Finn}\ and\ \citenamefont {Lizier}(2018)}]{finn2018pointwise}%
  \BibitemOpen
  \bibfield  {author} {\bibinfo {author} {\bibfnamefont {C.}~\bibnamefont {Finn}}\ and\ \bibinfo {author} {\bibfnamefont {J.~T.}\ \bibnamefont {Lizier}},\ }\href@noop {} {\bibfield  {journal} {\bibinfo  {journal} {Entropy}\ }\textbf {\bibinfo {volume} {20}},\ \bibinfo {pages} {297} (\bibinfo {year} {2018})}\BibitemShut {NoStop}%
\bibitem [{\citenamefont {Sparacino}\ \emph {et~al.}(2025{\natexlab{b}})\citenamefont {Sparacino}, \citenamefont {Antonacci}, \citenamefont {Mijatovic},\ and\ \citenamefont {Faes}}]{sparacino2025measuring}%
  \BibitemOpen
  \bibfield  {author} {\bibinfo {author} {\bibfnamefont {L.}~\bibnamefont {Sparacino}}, \bibinfo {author} {\bibfnamefont {Y.}~\bibnamefont {Antonacci}}, \bibinfo {author} {\bibfnamefont {G.}~\bibnamefont {Mijatovic}}, \ and\ \bibinfo {author} {\bibfnamefont {L.}~\bibnamefont {Faes}},\ }\href@noop {} {\bibfield  {journal} {\bibinfo  {journal} {Neurocomputing}\ }\textbf {\bibinfo {volume} {630}},\ \bibinfo {pages} {129675} (\bibinfo {year} {2025}{\natexlab{b}})}\BibitemShut {NoStop}%
\bibitem [{\citenamefont {Chicharro}(2011)}]{chicharro2011spectral}%
  \BibitemOpen
  \bibfield  {author} {\bibinfo {author} {\bibfnamefont {D.}~\bibnamefont {Chicharro}},\ }\href@noop {} {\bibfield  {journal} {\bibinfo  {journal} {Biological cybernetics}\ }\textbf {\bibinfo {volume} {105}},\ \bibinfo {pages} {331} (\bibinfo {year} {2011})}\BibitemShut {NoStop}%
\bibitem [{\citenamefont {Mijatovic}\ \emph {et~al.}(2025)\citenamefont {Mijatovic}, \citenamefont {Antonacci}, \citenamefont {Javorka}, \citenamefont {Marinazzo}, \citenamefont {Stramaglia},\ and\ \citenamefont {Faes}}]{mijatovic2025network}%
  \BibitemOpen
  \bibfield  {author} {\bibinfo {author} {\bibfnamefont {G.}~\bibnamefont {Mijatovic}}, \bibinfo {author} {\bibfnamefont {Y.}~\bibnamefont {Antonacci}}, \bibinfo {author} {\bibfnamefont {M.}~\bibnamefont {Javorka}}, \bibinfo {author} {\bibfnamefont {D.}~\bibnamefont {Marinazzo}}, \bibinfo {author} {\bibfnamefont {S.}~\bibnamefont {Stramaglia}}, \ and\ \bibinfo {author} {\bibfnamefont {L.}~\bibnamefont {Faes}},\ }\href@noop {} {\bibfield  {journal} {\bibinfo  {journal} {IEEE Transactions on Network Science and Engineering}\ } (\bibinfo {year} {2025})}\BibitemShut {NoStop}%
\bibitem [{\citenamefont {McPhaden}\ \emph {et~al.}(2006)\citenamefont {McPhaden}, \citenamefont {Zebiak},\ and\ \citenamefont {Glantz}}]{mcphaden2006enso}%
  \BibitemOpen
  \bibfield  {author} {\bibinfo {author} {\bibfnamefont {M.~J.}\ \bibnamefont {McPhaden}}, \bibinfo {author} {\bibfnamefont {S.~E.}\ \bibnamefont {Zebiak}}, \ and\ \bibinfo {author} {\bibfnamefont {M.~H.}\ \bibnamefont {Glantz}},\ }\href@noop {} {\bibfield  {journal} {\bibinfo  {journal} {science}\ }\textbf {\bibinfo {volume} {314}},\ \bibinfo {pages} {1740} (\bibinfo {year} {2006})}\BibitemShut {NoStop}%
\bibitem [{\citenamefont {Chang}\ \emph {et~al.}(2003)\citenamefont {Chang}, \citenamefont {Saravanan},\ and\ \citenamefont {Ji}}]{chang2003tropical}%
  \BibitemOpen
  \bibfield  {author} {\bibinfo {author} {\bibfnamefont {P.}~\bibnamefont {Chang}}, \bibinfo {author} {\bibfnamefont {R.}~\bibnamefont {Saravanan}}, \ and\ \bibinfo {author} {\bibfnamefont {L.}~\bibnamefont {Ji}},\ }\href@noop {} {\bibfield  {journal} {\bibinfo  {journal} {Geophysical Research Letters}\ }\textbf {\bibinfo {volume} {30}} (\bibinfo {year} {2003})}\BibitemShut {NoStop}%
\bibitem [{\citenamefont {Silini}\ \emph {et~al.}(2023)\citenamefont {Silini}, \citenamefont {Tirabassi}, \citenamefont {Barreiro}, \citenamefont {Ferranti},\ and\ \citenamefont {Masoller}}]{silini2023assessing}%
  \BibitemOpen
  \bibfield  {author} {\bibinfo {author} {\bibfnamefont {R.}~\bibnamefont {Silini}}, \bibinfo {author} {\bibfnamefont {G.}~\bibnamefont {Tirabassi}}, \bibinfo {author} {\bibfnamefont {M.}~\bibnamefont {Barreiro}}, \bibinfo {author} {\bibfnamefont {L.}~\bibnamefont {Ferranti}}, \ and\ \bibinfo {author} {\bibfnamefont {C.}~\bibnamefont {Masoller}},\ }\href@noop {} {\bibfield  {journal} {\bibinfo  {journal} {Climate Dynamics}\ }\textbf {\bibinfo {volume} {61}},\ \bibinfo {pages} {79} (\bibinfo {year} {2023})}\BibitemShut {NoStop}%
\bibitem [{\citenamefont {Stramaglia}\ \emph {et~al.}(2024)\citenamefont {Stramaglia}, \citenamefont {Faes}, \citenamefont {Cortes},\ and\ \citenamefont {Marinazzo}}]{stramaglia2024disentangling}%
  \BibitemOpen
  \bibfield  {author} {\bibinfo {author} {\bibfnamefont {S.}~\bibnamefont {Stramaglia}}, \bibinfo {author} {\bibfnamefont {L.}~\bibnamefont {Faes}}, \bibinfo {author} {\bibfnamefont {J.~M.}\ \bibnamefont {Cortes}}, \ and\ \bibinfo {author} {\bibfnamefont {D.}~\bibnamefont {Marinazzo}},\ }\href@noop {} {\bibfield  {journal} {\bibinfo  {journal} {Physical Review Research}\ }\textbf {\bibinfo {volume} {6}},\ \bibinfo {pages} {L032007} (\bibinfo {year} {2024})}\BibitemShut {NoStop}%
\bibitem [{\citenamefont {L{\"u}tkepohl}(2005)}]{lutkepohl2005new}%
  \BibitemOpen
  \bibfield  {author} {\bibinfo {author} {\bibfnamefont {H.}~\bibnamefont {L{\"u}tkepohl}},\ }\href@noop {} {\emph {\bibinfo {title} {New introduction to multiple time series analysis}}}\ (\bibinfo  {publisher} {Springer Science \& Business Media},\ \bibinfo {year} {2005})\BibitemShut {NoStop}%
\bibitem [{\citenamefont {Faes}\ \emph {et~al.}(2012)\citenamefont {Faes}, \citenamefont {Erla},\ and\ \citenamefont {Nollo}}]{faes2012measuring}%
  \BibitemOpen
  \bibfield  {author} {\bibinfo {author} {\bibfnamefont {L.}~\bibnamefont {Faes}}, \bibinfo {author} {\bibfnamefont {S.}~\bibnamefont {Erla}}, \ and\ \bibinfo {author} {\bibfnamefont {G.}~\bibnamefont {Nollo}},\ }\href@noop {} {\bibfield  {journal} {\bibinfo  {journal} {Computational and mathematical methods in medicine}\ }\textbf {\bibinfo {volume} {2012}},\ \bibinfo {pages} {140513} (\bibinfo {year} {2012})}\BibitemShut {NoStop}%
\bibitem [{\citenamefont {Hannan}(1979)}]{hannan1979statistical}%
  \BibitemOpen
  \bibfield  {author} {\bibinfo {author} {\bibfnamefont {E.~J.}\ \bibnamefont {Hannan}},\ }in\ \href@noop {} {\emph {\bibinfo {booktitle} {Developments in Statistics}}},\ Vol.~\bibinfo {volume} {2}\ (\bibinfo  {publisher} {Elsevier},\ \bibinfo {year} {1979})\ pp.\ \bibinfo {pages} {83--121}\BibitemShut {NoStop}%
\end{thebibliography}%

\end{document}